\documentclass[reprint,amsmath,amssymb,aps]{revtex4-2}

\usepackage{graphicx}
\usepackage{dcolumn}
\usepackage{bm}
\usepackage[dvipsnames]{xcolor}
\colorlet{LightRubineRed}{RubineRed!70!}
\colorlet{Mycolor1}{green!10!orange!90!}
\definecolor{Mycolor2}{HTML}{00F9DE}

\begin{document}

\preprint{APS/123-QED}

\title{Forward and inverse energy cascade and fluctuation relation in fluid turbulence adhere to Kolmogorov's refined similarity hypothesis}

\author{H.\,Yao}
\affiliation{
 Department of Mechanical Engineering \& Institute for Data Intensive Engineering \& Science, Johns Hopkins University
} 

\author{P.\,K.\,Yeung}
\affiliation{Department of Aerospace Engineering and Mechanical Engineering, Georgia Institute of Technology
}
\author{T.\,A.\,Zaki}
\author{C.\,Meneveau}\email{meneveau@jhu.edu}
 \affiliation{%
Mechanical Engineering \& IDIES, Johns Hopkins University
}%

\date{\today} 

\begin{abstract}

We study fluctuations of the local energy cascade rate $\Phi_\ell$ in turbulent flows at scales ($\ell$) in the inertial range.  According to the Kolmogorov refined similarity hypothesis (KRSH),  relevant statistical properties of $\Phi_\ell$ should depend on $\epsilon_\ell$, the viscous dissipation rate locally averaged over a sphere of size $\ell$, rather than on the global average dissipation. However, the validity of KRSH applied to $\Phi_\ell$ has not yet been tested from data.  Conditional averages such as $\langle \Phi_\ell|\epsilon_{\ell}\rangle$ as well as of higher-order moments are measured from Direct Numerical Simulations data, and results clearly adhere to the predictions from KRSH. Remarkably, the same is true  when considering forward ($\Phi_\ell>0$) and inverse ($\Phi_\ell<0$) cascade events separately. Measured ratios of forward and inverse cascade probability densities further show that a fluctuation relation adhering to the KRSH can be observed, raising the hope that important features of turbulence may be described using concepts from non-equilibrium thermodynamics.
\end{abstract}

\maketitle
 
The classic description of the energy cascade in turbulent flows states that the turbulent kinetic energy is extracted from large-scale eddies, transferred to smaller scale eddies, and finally dissipated into heat due to viscous friction \citep{richardson1922weather}. What is known from the Navier-Stokes equations is 
the celebrated $-4/5$ law  {\citep{kolmogorov1941local,frisch1995turbulence},
\textcolor{black}{
$
\langle \delta u_{\text{ll}}^3(r) \rangle \equiv \langle  ( [{\bf u}({\bf x}+{\bf r}) - {\bf u}({\bf x})]\cdot  {{\bf r} / r }  )^3 \rangle 
=- (4/5) \, r \, \langle \epsilon \rangle.
$
}
Here $\langle .. \rangle$ denotes statistical averaging, \textcolor{black}{$\delta u_{\text{ll}}(r)$} is the longitudinal velocity increment over distance $r$, $\epsilon$ is the viscous dissipation rate, and $r=|{\bf r}|$ is assumed to be inside the inertial range. The $4/5$ law means that in the inertial range, \textcolor{black}{$-(5/4)\langle \delta u_{\text{ll}}^3 \rangle / r$} can be interpreted as the energy {\color{black}cascade} rate and that the average direction {\color{black} of the cascade} is from large to small scales. However, it is well known that \textcolor{black}{$\delta u_{\text{ll}}$} and 
$\epsilon$
display strong intermittency  {\cite{kolmogorov1962refinement,meneveau1991multifractal,frisch1995turbulence}. To describe intermittency and anomalous scaling, Kolmogorov's second refined similarity hypothesis (KRSH) \cite{kolmogorov1962refinement} connects the statistics of 
$\delta u_{\text{ll}}(r)$
to 
the local   dissipation 
$\epsilon_r$, defined as the point-wise dissipation averaged in a ball of diameter $r$. 
KRSH has received strong support from early experimental measurements in which the dissipation $\epsilon_r$ had to be approximated by lower-dimensional data (e.g., \cite{stolovitzky1992kolmogorov,praskovsky1992experimental}) and also from later analyses based on 3D data, in which $\epsilon_r$ could be evaluated fully, from simulations \cite{wang1996examination,iyer2015refined,yeung2020advancing} 
or recent 3D experimental data \cite{lawson2019direct}. 

Most prior studies have started out with the KRSH formulated as a hypothesis inspired by dimensional analysis, but direct connections between KRSH and first-principles Navier-Stokes equations have often been lacking. \textcolor{black}{In this Letter, we revisit the equation of Hill {\citep{hill2002exact}}, from which a quantitative definition of the local cascade rate is possible. In this context we test the validity of the KRSH using conditional averaging based on local dissipation. We extend the  analysis and show new results concerning the probabilities of forward and inverse cascade rates. }
 
\textcolor{black}{The equation derived by} Hill  {\citep{hill2002exact}}, when written with no mean flow, for scales at which forcing can be neglected, and before averaging, reads:


\begin{equation}
\frac{\partial \delta u _i^2}{\partial t} + u^*_{j}\frac{\partial \delta u _i^2}{\partial x_j}  = 
-\frac{\partial \delta u _j\delta u _i^2}{\partial r_j}-\frac{8}{\rho}\frac{\partial p^*\delta u _i}{\partial r_i} + {\cal D} -  4\epsilon^*
\label{ins_KHMH_noint}
\end{equation}
where $\delta u_i = \delta u_i({\bf x};{\bf r}) =  u_i^+ - u_i^-$ is the velocity increment vector. 
The superscripts $+$ and $-$ represent two points ${\bf x}+{\bf r}/2$ and ${\bf x}-{\bf r}/2$ in the physical domain that have a separation vector $r_i = x^+_i - x^-_i$ and middle point 
$x_i = (x^+_i + x^-_i)/2$. The superscript $*$ denotes the average value between two points
, e.g., the average dissipation is defined as $\epsilon^* = (\epsilon^+ +\epsilon^-)/2$. In this paper $\epsilon$ denotes the ``pseudo-dissipation'', defined as $\epsilon =\nu ({\partial u_i}/{\partial x_j})^2$ and ${\cal D}$ is a (small) viscous term. Many variants of this equation 
\textcolor{black}{(also called Karman-Howarth-Monin-Hill, KHMH, equation in \citep{yasuda2018spatio}) have been studied \cite{Monin_Yaglom_1975,danaila_anselmet_zhou_antonia_2001}.}
To 
make connection to the RKSH and $\epsilon_\ell$ at some scale $r=\ell$, Eq.\,\ref{ins_KHMH_noint} at any point ${\bf x}$ can be integrated over a sphere in scale {\bf r}-space up to a diameter $r=\ell$ (radius $\ell/2$). The resulting equation describes the evolution of local kinetic energy up to scale $\ell$ defined as $k_\ell=
(1/2 \Omega_\ell) \int_{\Omega_{\ell}} \frac{1}{2} \delta u _i^2 d^3{\bf r}_s$ 
with the radius vector ${\bf r}_s = {\bf r}/2$ integrated up  to $\ell/2$.
 When divided by the volume of the sphere $(\Omega_\ell=\frac{4}{3}\pi( {\ell}/{2})^3)$ and a factor of 4, the locally integrated form of Eq. \ref{ins_KHMH_noint} becomes
\begin{equation}
  \frac{\tilde{d} k_\ell}{d t}   =  \Phi_\ell  +   P_\ell  +  D_\ell  - \epsilon_\ell,  
 \label{KHMH_local}
\end{equation}
where $ \epsilon_\ell({\bf x}) \equiv \frac{1}{\Omega_\ell}\int_{\Omega_{\ell}}
 \epsilon^*({\bf x},{\bf r})\,d^3{\bf r}_s $
is the $\ell$-averaged 
dissipation envisioned in the RKSH,
and 
\begin{equation}
\Phi_\ell({\bf x})  \equiv -\frac{3}{4\,\ell}\frac{1}{S_\ell}\oint\limits_{S_{\ell}} \delta u _i^2\,\delta u _j\,  \hat{n}_j dS 
\end{equation}
is interpreted as the local energy cascade rate. Note that Gauss theorem is used to evaluate the first term on the RHS of Eq.\,\ref{ins_KHMH_noint} as an integral over the $r$-sphere's surface, with area element $\hat {n}_j dS$, with $\hat {\bf n} = {\bf r}/|{\bf r}|$, and $S_\ell = 4\pi (\ell/2)^2$.
Eq.\,\ref{KHMH_local} also includes ${{\tilde d} k_\ell}/{d t}$, the 
advective rate of change of kinetic energy $k_\ell$ defined as 
$  {{\tilde d} k_\ell}/{d t} \equiv 
(2 \,\Omega_\ell)^{-1}\iiint_{\Omega_{\ell}} \left(\partial_t \delta u _i^2/2 + u^*_{j}  \partial_j \delta u _i^2/2 \right) d^3{\bf r}_s $,  and
$
P_\ell \equiv -6 \, (\ell S_\ell)^{-1}  \oint_{S_{\ell}} \, (p^*/\rho) \, \delta u _j \, \hat{n}_j \,dS
$, 
a surface averaged pressure work term at scale $\ell$. The term $D_\ell = \int_{\Omega_{\ell}}{\cal D} d^3{\bf r}_s /4 =  \nu/(8 \Omega_{\ell})\int_{\Omega_{\ell}} [\partial^2 \delta u^2_i/\partial x_j^2 + 4 \, \partial^2 \delta u^2_i / \partial r_j^2 ] d^3{\bf r}_s$ represents viscous diffusion of $k_\ell$ both in position and scale space. We consider $\ell$ to be in the inertial range, therefore $D_\ell$ is negligible. Eq. \ref{KHMH_local}, \textcolor{black}{the ``scale-integrated local KH'' (Kolmogorov-Hill) equation} is local, valid at any {\color{black} $({\bf x},t)$ position and time (see also \cite{Duchon_Robert_2000,eyink2002local,dubrulle2019beyond}}).

A reformulation of the KRSH in the present context is that the statistics of $\Phi_\ell$ 
only depend on the statistics of $\epsilon_\ell$ (i.e.\,that $\Phi_\ell=V_{\Phi} \epsilon_\ell$ with random variable $V_{\Phi}$ independent of $\ell$ and $\epsilon_\ell$) in the inertial range. In particular, the conditional average of the cascade rate should obey $\langle \Phi_\ell | \epsilon_\ell\rangle = \epsilon_\ell$ \textcolor{black}{and $\langle V_\Phi\rangle=1$}. In fact, referring back to  
Eq.\,\ref{KHMH_local}, we may take its conditional average and write (neglecting $D_\ell$, and using $\langle\epsilon_\ell|\epsilon_\ell\rangle =\epsilon_\ell$) 
\begin{equation}
 \langle {\tilde{d} k_\ell}/{d t} \,| \epsilon_\ell \rangle = \langle \Phi_\ell |\epsilon_\ell\rangle + \langle P_\ell |\epsilon_\ell\rangle - \epsilon_\ell.
 \label{KHMH_epsilon}
\end{equation}
Hence, a consequence of the KRSH with 
Eq. \ref{KHMH_local}
is that the conditional average of $W_\ell
\equiv \widetilde{d}k_\ell/{d t} -  P_\ell$  must vanish, i.e.\,$\langle W_\ell |\epsilon_r\rangle = 0$
(or both $\langle \widetilde{d}k_\ell/{d t}|\epsilon_\ell \rangle=0$ and  $\langle P_\ell|\epsilon_\ell \rangle=0$).


Prior measurements of $\Phi_\ell$ \cite{Yao2023comparing,yao2023entropy} 
show that it can be both positive and negative locally, and {\color{black} moreover \cite{yao2023entropy} suggests that} the ratio $\Psi_\ell = \Phi_\ell/k_\ell$ can be understood as an entropy generation (or phase-space contraction) rate,  where $k_\ell$ is interpreted as the ``temperature of turbulence'' \cite{yao2023entropy}.  \textcolor{black} {There is growing interest in connecting turbulence with thermodynamics concepts, e.g., focusing on model systems \cite{chorin1991equilibrium,paladin1987anomalous}, and on possible definitions of entropy \cite{vela2021entropy} and temperature \cite{castaing1996temperature}.} A prediction about entropy generation rates in  non-equilibrium thermodynamics is the ``Fluctuation Relation'' (FR) \cite{evans1993probability,gallavotti1995dynamical}
. 
\textcolor{black} {Prior authors have examined the FR, most often focusing on stochastic models \cite{fuchs2020small} but also on fluctuations in global power input \cite{zonta2016entropy} and spectral energy transfer \cite{porporato2020fluctuation}. As in \cite{yao2023entropy} we consider intermittent fluctuations of spatially local quantities instead.}
When written for turbulence entropy generation rates the FR states that the ratio of probability densities of positive and negative $\Psi_\ell$ follows the exponential behavior
$P(\Psi_\ell)/P(-\Psi_\ell)=\exp(\Psi_\ell \tau_\ell)$, where $\tau_\ell$ is a characteristic time-scale. The recent results of \textcolor{black}{\cite{yao2023entropy}} confirm the FR relationship for isotropic turbulence.
However, in the prior analysis \cite{yao2023entropy}} the time-scale 
was defined using the average dissipation, \textcolor{black}{ i.e. $\overline{\tau}_\ell = \langle \epsilon\rangle^{-1/3}\ell^{2/3}$}. 
In other words, it did not take into account effects of intermittency in which different regions of the flow with different $\epsilon_\ell$ values could behave differently.  

\textcolor{black}{The specific aims in this Letter are to} investigate whether data support the KRSH in the context of the dynamics of turbulent kinetic energy at scale $\ell$ as described by 
Eq. \ref{KHMH_local}, 
i.e., whether predictions from KRSH hold for (i) conditional moments of $\Phi_\ell$, (ii) for the combined unsteady and pressure terms $W_\ell$, and (iii) for positive and negative cascade rates as well as for the fluctuation relation from non-equilibrium thermodynamics. 

We evaluate terms in 
Eq. \ref{KHMH_local}
using data from Direct Numerical Simulation of isotropic turbulence at $R_\lambda \approx $1,250 
(data obtained from the JHTDB database \cite{li2008public,yeung2015extreme}). 
Surface averages needed to evaluate $\Phi_\ell$ are measured by discretizing the outer surface of diameter $\ell$ into 500 point pairs ($+$ and $-$ points) that are approximately uniformly distributed on the sphere. Velocities for $\delta u_i$ are downloaded from JHTDB. 
Volume integrals $\epsilon_{\ell}$ and $k_{\ell}$ are evaluated similarly by integrating over five concentric spheres. The accuracy of this method of integration has been tested by increasing the number of points used in the discretization and confirming indistinguishable results are obtained. For $\epsilon$ we use JHTDB's GetVelocityGradient method with 4th-order centered finite differencing. Taking dissipation as an example, panel (a) in figure \ref{Diss_ins_int} shows point-wise normalized dissipation $\epsilon({\bf x})/\langle \epsilon \rangle$ computed on a planar cut across the data (the plane shown corresponds to $500\times 500$ grid points). Panel (b) of figure \ref{Diss_ins_int} shows a sphere with diameter $\ell$.  

\begin{figure}
 \centering
    \includegraphics[scale=0.237]{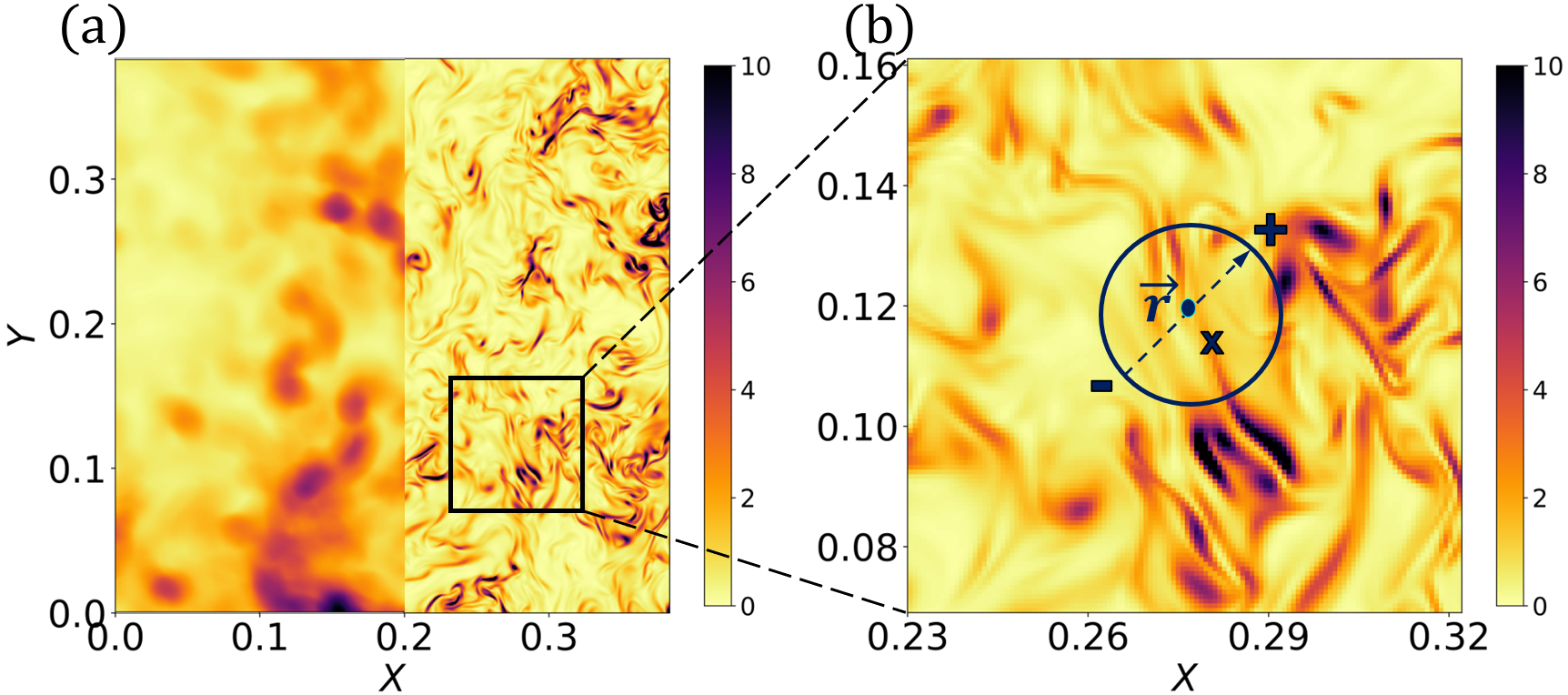}
    \vspace{-0.25in}
    \caption{ (a) Spatial distribution of local dissipation rate normalized by $\langle \epsilon\rangle$ on a plane in a small subset of isotropic turbulence at $R_\lambda=$1,250. The left portion shows $\epsilon_\ell$ distribution obtained from spherical filtering. (b) Zoomed-in portion of panel (a) also showing a sphere  with a diameter $\ell=45 \eta$ marked as the black circle. The black dash arrow represents ${\bf r}$ separating the two points $+$ and $-$. 
    }
    \label{Diss_ins_int}
\end{figure}

\begin{figure}
\centering
\includegraphics[scale=0.32]{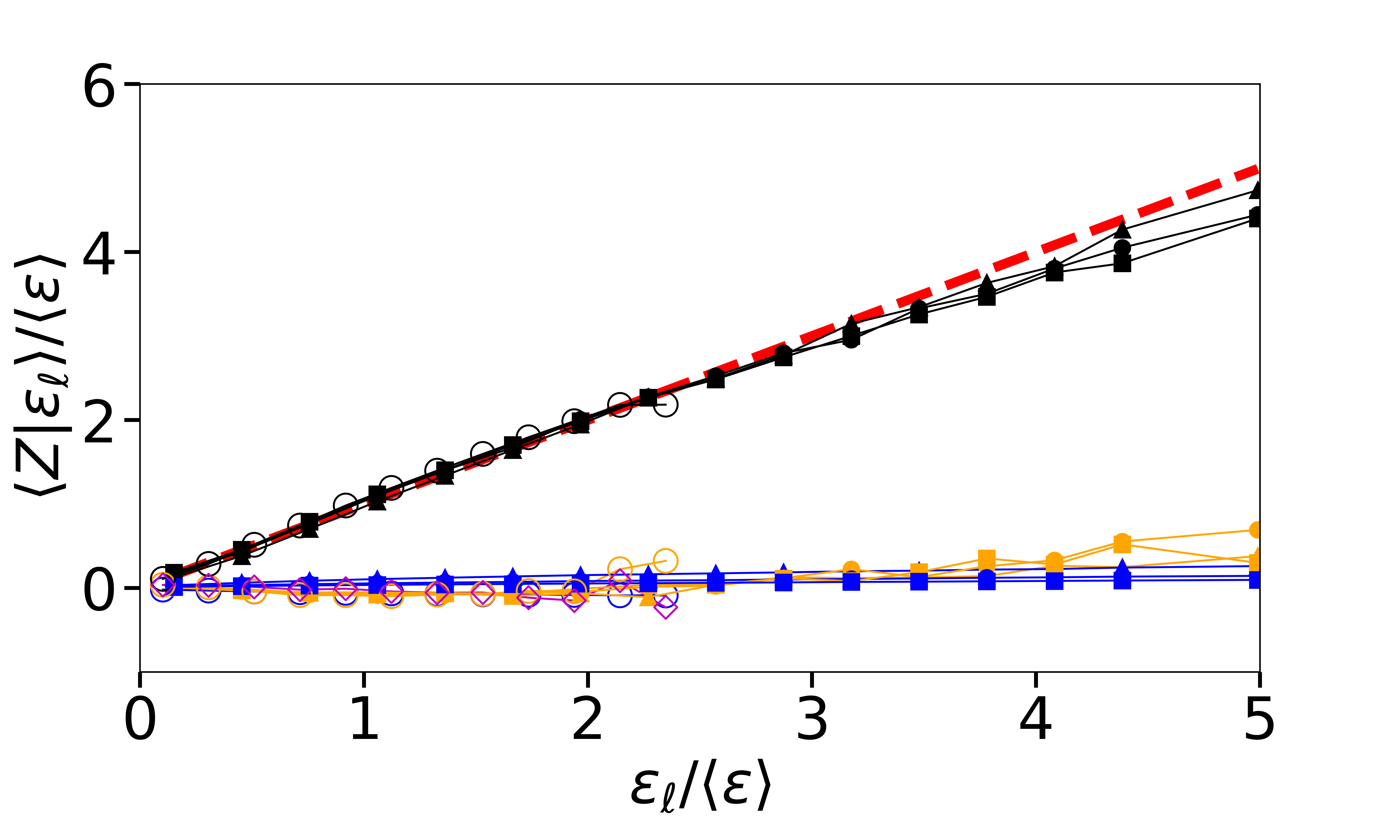}
   \vspace{-0.15in}
    \caption{Conditional averages of terms in the \textcolor{black}{scale-integrated local KH equation} 
    based on 
    $\epsilon_\ell$, i.e., $Z = \Phi_\ell$ (black symbols and lines), $Z = P_\ell$ (yellow symbols and lines), $Z = D_\ell$ (blue symbols and lines). The red \textcolor{black}{dashed}  line indicates $\epsilon_\ell$.  Different symbols denote different scales  $\ell/L = 0.012 $ (triangles, \textcolor{black}{$(30 \eta)$}), $0.018$ (circles, $(45 \eta)$) and $0.024$ (squares,  \textcolor{black}{$(60 \eta)$}). Solid symbols: Data  from  DNS 
    at $R_\lambda=$1,250 (solid symbols) and 
    $R_\lambda=$430 at $\ell/L = 0.092 \, (45 \eta)$ (open circles), for which 
    $Z = \widetilde{d} k_\ell/d t$ (purple diamonds) is included. 
    }
    \label{FPV_epr_budget304560_8192}
\end{figure}

Figure \ref{FPV_epr_budget304560_8192} shows the 
conditional average
$\langle \Phi_\ell | \epsilon_\ell\rangle$ as function of $\epsilon_\ell$. Statistics are computed using 2,000,000 randomly distributed spheres across the entire $8192^3$ isotropic turbulence dataset (isotropic8192) \citep{yeung2015extreme}. The analysis considers two length scales in the inertial range and one approaching the viscous range, namely $\ell=0.024 L = 60 \eta$, $\ell=0.018 L = 45\eta$, and $\ell=0.012 L = 30\eta$, respectively, where $L = 1.24$ is the integral scale and the Kolmogorov scale is $\eta = (\nu^3/\langle \epsilon \rangle)^{1/4} = 4.98 \times 10^{-4}$.  
It is 
apparent that the dominant terms are $\langle \Phi_\ell |\epsilon_\ell\rangle$ (black dots) and $\epsilon_\ell$ itself (red dash line with unit slope). These are equal for most of the  range of $\epsilon_\ell$ for which reliable statistics can be collected. The good collapse  $\langle \Phi_\ell |\epsilon_\ell\rangle  \approx \epsilon_\ell$   provides 
{\color{black} clear}
support for the KRSH in the context of terms appearing in 
Eq. \ref{KHMH_local}.
Also plotted in Fig.\,\ref{FPV_epr_budget304560_8192} are the conditional averages of the pressure term, $\langle P_\ell | \epsilon_\ell\rangle$, and the viscous term, $\langle D_\ell | \epsilon_\ell\rangle$. We can see  that the contribution of the pressure term (yellow squares) is negligible at all three length scales over most of the range. The viscous term (solid blue lines) is also negligibly small as expected. Approaching the largest values of $\epsilon_\ell/\langle \epsilon \rangle$ we observe saturation of $\Phi_\ell$ that is compensated by small rise of the pressure term  
(same results are obtained when using the full viscous dissipation $\nu(\partial u_i\partial x_j)(\partial u_i\partial x_j+\partial u_i\partial x_j)$ instead of the pseudo-dissipation $\nu(\partial u_i\partial x_j)^2$).

Moreover, based on Eq.\,\ref{KHMH_epsilon}, we can conclude that $\langle \widetilde{d} k_\ell/d t|\epsilon_\ell \rangle \approx 0$, given that $\langle P_\ell |\epsilon_\ell\rangle \approx 0$, $\langle D_\ell |\epsilon_\ell\rangle \approx 0$, and $\langle \Phi_\ell |\epsilon_\ell\rangle \approx \epsilon_\ell$. To verify this result via explicit measurement, we computed the terms in Eq.\,\ref{KHMH_epsilon} using the   isotropic1024 \citep{li2008public} dataset. It has a smaller size of $1024^3$ grid points and a lower Reynolds number $R_\lambda = 430$, but includes temporally consecutive snapshots allowing us to calculate time derivatives. Fig.\, \ref{FPV_epr_budget304560_8192} (open symbols)
shows that $\langle \Phi_\ell |\epsilon_\ell\rangle \approx \epsilon_\ell$ still holds very well at this lower Reynolds number, that the pressure and viscous terms are again close to zero and that $\langle \widetilde{d} k_\ell/d t|\epsilon_\ell \rangle \approx 0$.  We conclude that the data provide strong direct support to the KRSH relating $\Phi_\ell$ and $\epsilon_\ell$ and that the 
pressure and unsteadiness terms vanish in the inertial range. 


A further implication of KRSH relates to higher order moments. \textcolor{black}{It implies that  
$\langle \Phi_\ell^q | \epsilon_\ell\rangle  = \langle V_\Phi^q\rangle  \, \epsilon_\ell^q$}. In the inertial range, since $\Phi_\ell=\epsilon_\ell+W_\ell$ locally and instantaneously, raising to the $q$-power, expanding, and taking the conditional average yields
$\langle \Phi_\ell^q |\epsilon_\ell\rangle  = \sum_{n=0}^q  \binom{q}{n}  \,\,\epsilon_\ell^{q-n} \,\,\langle W_\ell^{n} |\epsilon_\ell\rangle.
$
 Thus, for KRSH to hold (i.e.\,for 
 $\langle \Phi_\ell^q | \epsilon_\ell\rangle   \propto \epsilon_\ell^q$)
 the conditional moments of $W_\ell$ must follow the same behavior, i.e., $\langle W_\ell^n | \epsilon_\ell\rangle \propto \epsilon_\ell^n$. Both the KRSH prediction for $\langle \Phi_\ell^q | \epsilon_\ell\rangle$ and $\langle W_\ell^n | \epsilon_\ell\rangle$ can be tested by measuring and plotting $\langle \Phi_\ell^q | \epsilon_\ell\rangle^{1/q}$ and $\langle W_\ell^n | \epsilon_\ell\rangle^{1/n}$ as function of $\epsilon_\ell$ and
testing for linear behavior. Results are shown in Fig.\,\ref{FPV_epr_moments} for $q=2,3$ and $n=2,3$. Clearly, the proportionality holds, with linear trends visible for {\color{black} these} moment orders over the range of dissipation values. 
 
\begin{figure}
 \centering
    \includegraphics[scale=0.235]{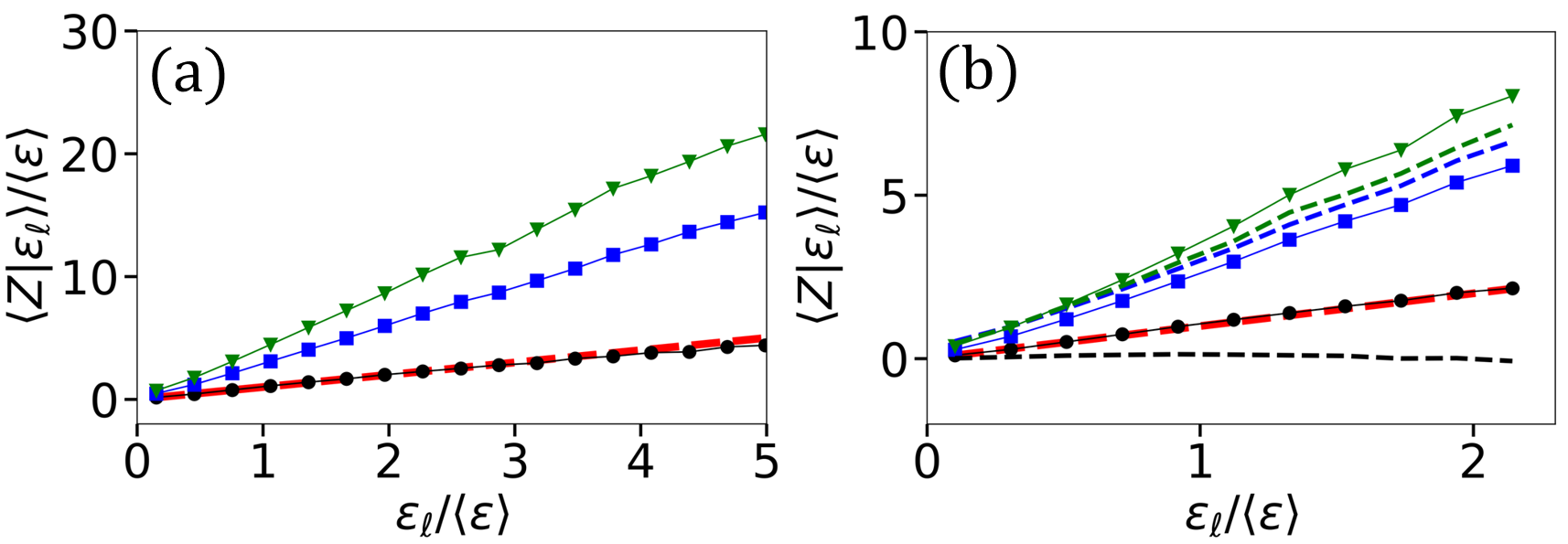}
     \vspace{-0.15in}
    \caption{Conditional averaged $Z = \Phi^q_\ell$ (symbols) for the isotropic8192 (a) and isotropic1024 (b) datasets, and $Z = W^n_\ell$ (dashed lines for the  isotropic1024  dataset), for $\ell=45 \eta$, plotted as function of the conditioning variable $\epsilon_\ell$. Results for $q,n=1,2,3$ are shown in black, blue, and green respectively.  All terms 
    display linear trends with $\epsilon_\ell$, consistent with the KRSH.}
    \label{FPV_epr_moments}
\end{figure}

We now turn to further consequences of the KRSH that are directly related to the direction of the energy cascade, i.e. we examine if KRSH may be applicable even to those regions of the flow where $\Phi_\ell <0$, i.e., those displaying 
 only inverse cascading, or $\Phi_\ell >0$, i.e, those displaying only forward cascading.
An implication of KRSH is that the conditional average of only positive and only negative values of $\Phi_\ell$ should also be proportional to $\epsilon_\ell$.
To investigate this prediction, we split the samples of $\Phi_\ell$ by its sign and perform conditional averaging based on $\epsilon_\ell$.  We first observe that there are about twice as many samples with $\Phi_\ell>0$ than with $\Phi_\ell<0$, specifically, if ${\rm Pr}(\Phi_\ell>0|\epsilon_\ell)$ and ${\rm Pr}(\Phi_\ell<0|\epsilon_\ell)$ are the conditional total probabilities associated with signs of $\Phi_\ell$ in any $\epsilon_\ell$ bin,  we measure ${\rm Pr}(\Phi_\ell>0|\epsilon_\ell)/{\rm Pr}(\Phi_\ell<0|\epsilon_\ell) \approx 2.0$ (see blue line in Fig.\,\ref{F_epr_FNFr}).
In the inertial range, the ratio is approximately 2.0 (consistent with results from Ref.\,\cite{cardesa2017turbulent}, who defined 
energy cascade rate by detailed evaluations of intersection of eddies and lifetimes). 
From normalization we conclude that 
${\rm Pr}(\Phi_\ell>0|\epsilon_\ell) \approx 2/3$ and  
${\rm Pr}(\Phi_\ell<0|\epsilon_\ell) \approx 1/3$.  And since 
$\langle \Phi_\ell | \epsilon_\ell \rangle = \langle \Phi_\ell | \epsilon_\ell, \Phi_\ell >0 \rangle {\rm Pr}(\Phi_\ell>0|\epsilon_\ell) + \langle \Phi_\ell | \epsilon_\ell, \Phi_\ell <0 \rangle {\rm Pr}(\Phi_\ell<0|\epsilon_\ell)$ and the data already showed $\langle \Phi_\ell | \epsilon_\ell \rangle \approx \epsilon_\ell$, the KRSH further implies that $\langle \Phi_\ell | \epsilon_\ell, \Phi_\ell < 0 \rangle \approx  3 \, \epsilon_\ell - 2 \, \langle \Phi_\ell | \epsilon_\ell, \Phi_\ell > 0 \rangle$. These predictions from RKSH are  tested 
 in Fig.\,\ref{F_epr_FNFr}, showing that 
 $\langle \Phi_\ell | \epsilon_\ell, \Phi_\ell > 0 \rangle \approx 2 \, \epsilon_\ell$ and $\langle \Phi_\ell | \epsilon_\ell, \Phi_\ell < 0 \rangle \approx -1 \, \epsilon_\ell$. Clearly KRSH holds even for the positive and negative regions separately. For completeness, we also show the conditional average of the traditional third-order longitudinal structure function, which under the assumption of isotropy and KRSH follows 
$- (5/4 \ell) \langle (\delta u_j \hat{n}_j)^3  | \epsilon_\ell\rangle = \epsilon_\ell$.

\begin{figure}
 \centering
     \includegraphics[scale=0.25]
     {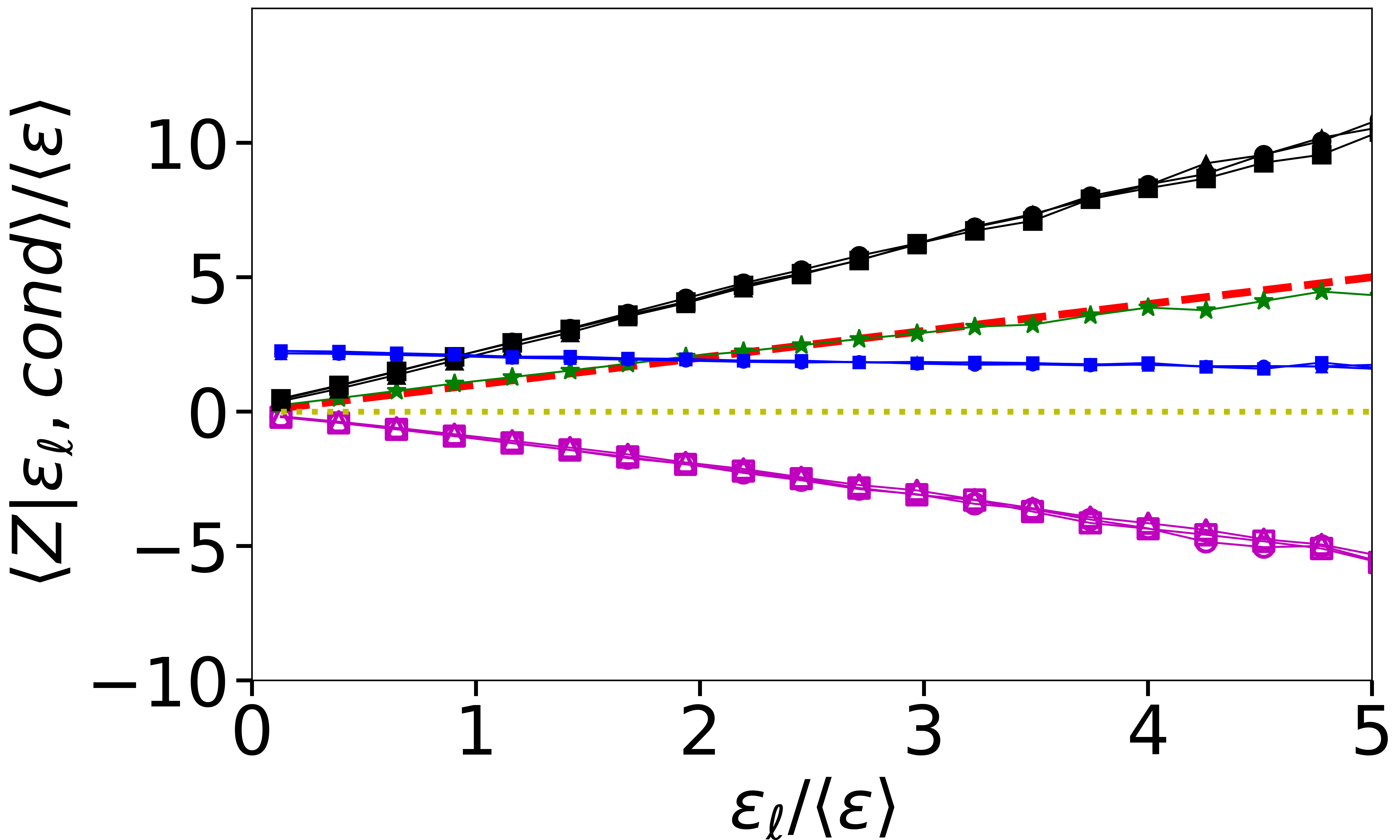}
      \vspace{-0.15in}
    \caption{Conditionally averaged positive (forward cascade, full symbols, $Z=\Phi_\ell$, cond: $\Phi_\ell>0$) and negative (inverse cascade, empty symbols, $Z=\Phi_\ell$, cond: $\Phi_\ell<0$)
    for the isotropic8192 dataset, and for three scales $\ell$ (symbols are the same as in Fig. 2). The green stars show the standard 4/5-law quantity based on longitudinal structure function, i.e., for $Z=(-5/4\ell) (\delta u_j \hat{n}_j)^3$, and no {\color{black} additional condition beyond $\epsilon_\ell$}. Blue symbols is
${\rm Pr}(\Phi_\ell>0|\epsilon_\ell)/{\rm Pr}(\Phi_\ell<0|\epsilon_\ell)$ ($\approx 2.0$). 
    }
    \label{F_epr_FNFr}
\end{figure}

Next, following Ref. \textcolor{black}{\cite{yao2023entropy}} we examine the fluctuation relation from non-equilibrium thermodynamics, but \textcolor{black} {instead of using the overall dissipation rate to define the characteristic eddy turn-over timescale, we here use conditioning} on various values of local dissipation $\epsilon_\ell$. Figure \ref{fig:condpdfpsi}(a) shows the conditional PDF $P(\Psi_\ell|\epsilon_\ell)$ of the entropy generation rate $\Psi_\ell = \Phi_\ell/k_\ell$ for $\ell/\eta=45$, conditioned for various values of $\epsilon_\ell$ ranging from $\epsilon_\ell/\langle \epsilon \rangle = 0.15$ to 4.2. As in \textcolor{black}{\cite{yao2023entropy}}, exponential tails are found, with steeper slopes on the negative side than on the positive one, and approximately twice as steep. Remarkably, when multiplying $\Psi_\ell$ by the corresponding \textcolor{black}{local} turn-over time-scale $\tau_\ell = \epsilon_\ell^{-1/3} \ell^{2/3}$ where  $\epsilon_\ell$ is the value used to bin the data, excellent collapse is observed, see Fig. \ref{fig:condpdfpsi} (b). If the PDFs are approximated as pure exponentials, with slope \textcolor{black}{magnitudes} $\alpha_-$ for $\Psi_\ell<0$ and $\alpha_+$ for $\Psi_\ell>0$, 
it is evident from Fig. \ref{fig:condpdfpsi}(b) that $\alpha_+ \approx 1$ and $\alpha_- \approx 2$. For such two-sided exponential PDFs, it is easy to show that 
${\rm Pr}(\Psi_\ell<0)/{\rm Pr}(\Psi_\ell>0)=\alpha_+/\alpha_-$, consistent with the 1:2 ratio discussed above, independent of $\epsilon_\ell$. 
\begin{figure}
 \centering
     \includegraphics[scale=0.239]
     {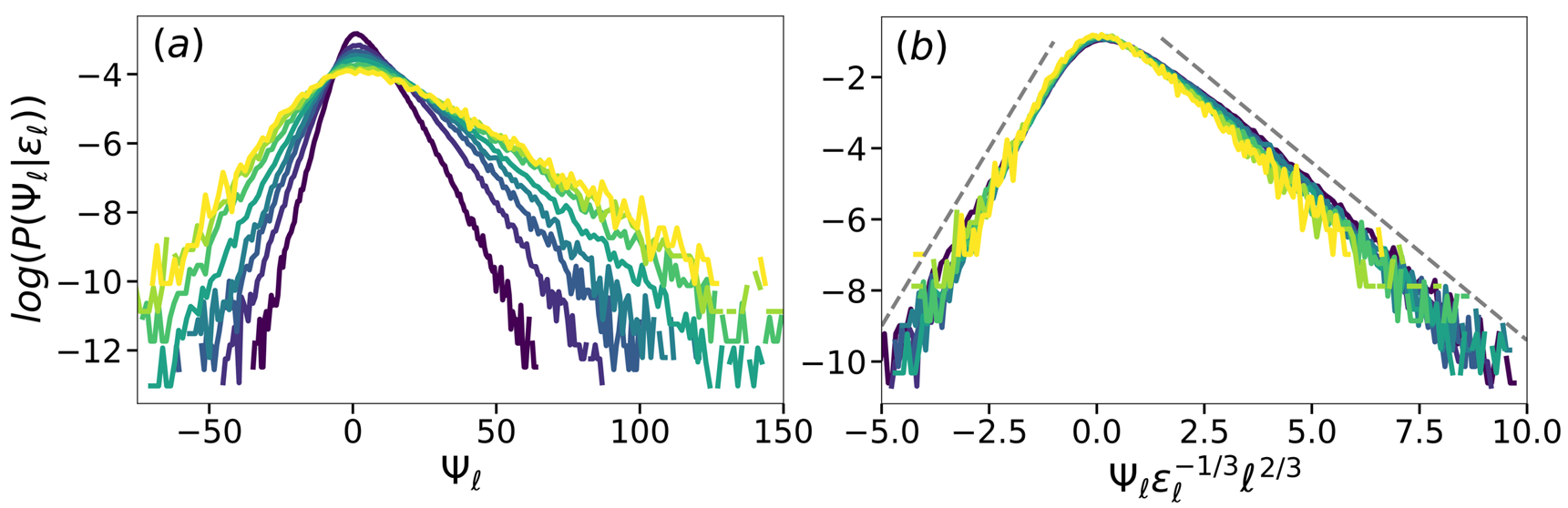}
      \vspace{-0.15in}
    \caption{Conditional PDFs of $\Psi_\ell$, for $\ell=45\eta$ conditioned on ranges in bins centered at $\epsilon_\ell / \langle \epsilon \rangle = 0.15$ (black), 0.45, 0.75, 1.05, 1.8, 3.0, 4.2, and 5.4 (yellow). The grey dash lines have slopes = $2$ (left) and $-1$ (right). Natural logarithm is used. 
    }
    \label{fig:condpdfpsi}
\end{figure}
 Finally, the FR can be tested by plotting $\log[P(\Psi_\ell|\epsilon_\ell)/P(-\Psi_\ell|\epsilon_\ell)]$  versus $\Psi_\ell \tau_\ell$ (see Fig. \ref{fig:condFR}). The result shows good collapse and an approximately linear trend (especially at $\Psi_\ell \tau_\ell>1$, thus 
 \textcolor{black}{providing empirical/approximate support for} the fluctuation relation for turbulence even when conditioning on different values of $\epsilon_\ell$, and using $\epsilon_\ell$ to establish the relevant turn-over time-scale.  For the exponential approximation of the conditional PDFs, the slope in the FR plot is simply $\alpha_- - \alpha_+$, which is nearly unity (as observed originally in \textcolor{black}{\cite{yao2023entropy}}), and quite consistent with $\alpha_- \approx 2$ while $\alpha_+ \approx 1$.

\begin{figure}
 \centering
     \includegraphics[scale=0.25]
     {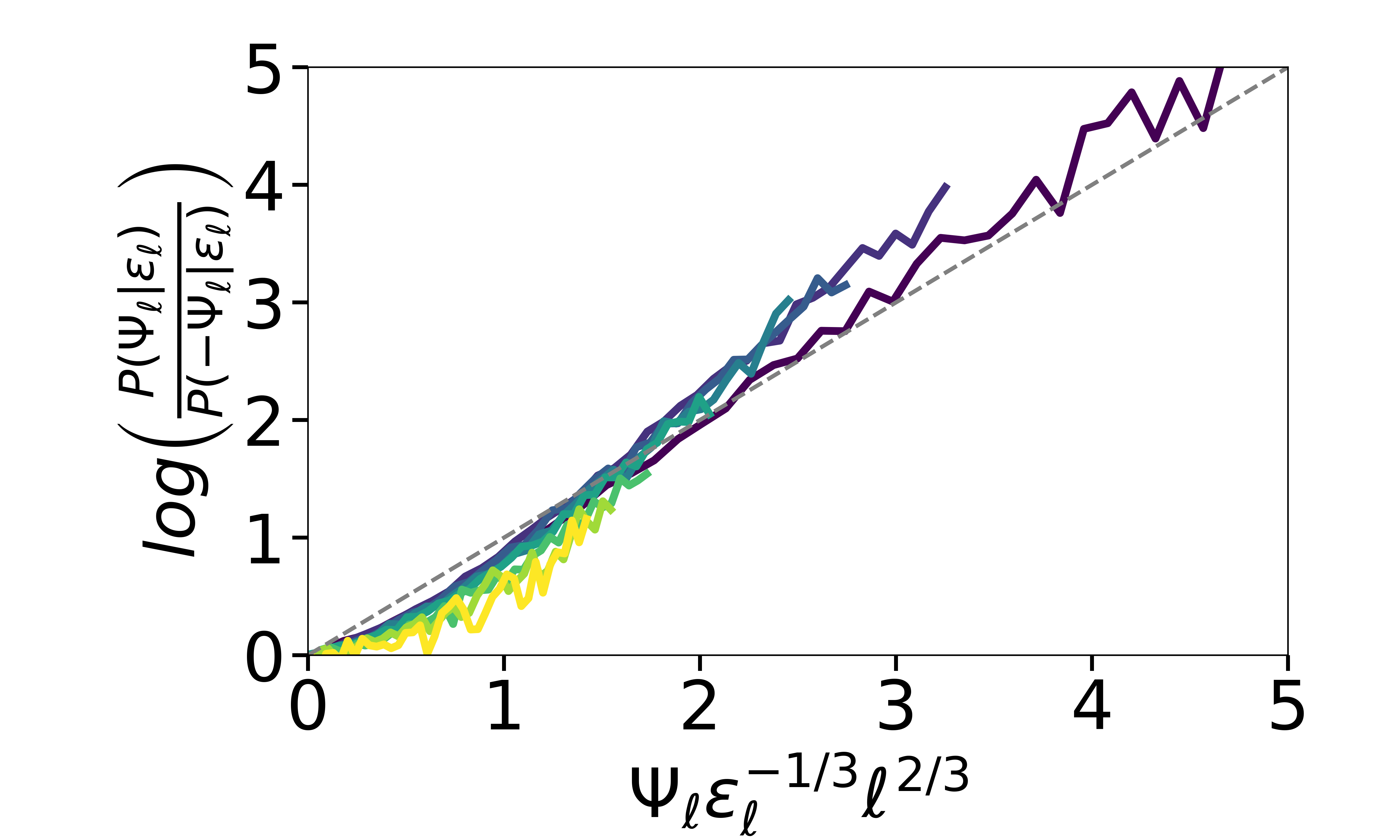}
      \vspace{-0.15in}
    \caption{Test of conditional fluctuation relation for different values of $\epsilon_\ell / \langle \epsilon \rangle$ (same line colors as in Fig. \ref{fig:condpdfpsi}). The grey line has slope $=1$. 
    }
    \label{fig:condFR}
\end{figure}

In summary, we examined the KRSH involving the  
\textcolor{black}{local}
dissipation $\epsilon_\ell$ in the context of an equation (\textcolor{black}{Eq. \ref{KHMH_local}}) derived exactly from the Navier-Stokes equations.
 Results from two DNS of forced isotropic turbulence provided strong support for the validity of KRSH for the cascade rate $\Phi_\ell$, its moments, and also for moments of other terms appearing in the dynamical equation.
Furthermore the data support a strong version of the KRSH when positive and negative cascade rates are considered separately, each of which scale proportional to $\epsilon_\ell$. Finally, the fluctuation relation from non-equilibrium thermodynamics is shown to hold to a good approximation, relating the probability of forward and backward cascade events using the local time-scale based on $\epsilon_\ell$ and $\ell$. \textcolor{black}{This result represents the first time that the fluctuation relation is observed in the inertial range of 3D turbulence (not model systems), based on conditional averages as in the KRSH.}

Present \textcolor{black}{findings} connecting KRSH directly to a dynamical equation derived from the Navier-Stokes equations
as well as to basic principles of non-equilibrium thermodynamics could help in developing improved theories and models of the turbulence cascade process. Additional questions arise, such as what are the effects of time-averaging in defining the entropy generation rate $\Psi_\ell$ \textcolor{black}{ (we have applied the FR at some turn-over time $t=\tau_\ell$ and tested FR only as far as the fluctuations in $\Psi_\ell$ are concerned)}, what occurs when $\ell$ approaches limits of the inertial range, and do the results hold for flows with mean shear and walls? \textcolor{black}{Finally, it bears recalling that 
the FR is applicable to systems in which the microscopic dynamics are time-reversible (but, see discussion in \cite{gallavotti2020ensembles}). As argued in \cite{yao2023entropy}, we here regard the microscopic degrees of freedom to be the inertial-range eddies smaller than $\ell$ whose evolution is governed by nearly inviscid, thus time-reversible, dynamics at least for some small finite time. While the actual turbulent state we are analyzing is already far from equilibrium so that the overall statistics of velocity increments (and $\Phi_\ell$, $\Psi_\ell$) display the established asymmetry, the actual dynamics in the inertial range can be time-reversed for some time \cite{vela2021entropy}. Still, significant questions remain about 
connections between non-equilibrium thermodynamics and 3D turbulence in the inertial range. } 

{\bf Acknowledgements:} We thank R. Hill for comments on an early draft of this work.  The work is supported by NSF (CSSI-2103874) and the contributions from the JHTDB team are gratefully acknowledged. 

\bibliographystyle{apsrev4-2}
\bibliography{YaoYeungZakiMeneveau_PRL_2023} 

\begin{thebibliography}{33}%
\makeatletter
\providecommand \@ifxundefined [1]{%
 \@ifx{#1\undefined}
}%
\providecommand \@ifnum [1]{%
 \ifnum #1\expandafter \@firstoftwo
 \else \expandafter \@secondoftwo
 \fi
}%
\providecommand \@ifx [1]{%
 \ifx #1\expandafter \@firstoftwo
 \else \expandafter \@secondoftwo
 \fi
}%
\providecommand \natexlab [1]{#1}%
\providecommand \enquote  [1]{``#1''}%
\providecommand \bibnamefont  [1]{#1}%
\providecommand \bibfnamefont [1]{#1}%
\providecommand \citenamefont [1]{#1}%
\providecommand \href@noop [0]{\@secondoftwo}%
\providecommand \href [0]{\begingroup \@sanitize@url \@href}%
\providecommand \@href[1]{\@@startlink{#1}\@@href}%
\providecommand \@@href[1]{\endgroup#1\@@endlink}%
\providecommand \@sanitize@url [0]{\catcode `\\12\catcode `\$12\catcode `\&12\catcode `\#12\catcode `\^12\catcode `\_12\catcode `\%12\relax}%
\providecommand \@@startlink[1]{}%
\providecommand \@@endlink[0]{}%
\providecommand \url  [0]{\begingroup\@sanitize@url \@url }%
\providecommand \@url [1]{\endgroup\@href {#1}{\urlprefix }}%
\providecommand \urlprefix  [0]{URL }%
\providecommand \Eprint [0]{\href }%
\providecommand \doibase [0]{https://doi.org/}%
\providecommand \selectlanguage [0]{\@gobble}%
\providecommand \bibinfo  [0]{\@secondoftwo}%
\providecommand \bibfield  [0]{\@secondoftwo}%
\providecommand \translation [1]{[#1]}%
\providecommand \BibitemOpen [0]{}%
\providecommand \bibitemStop [0]{}%
\providecommand \bibitemNoStop [0]{.\EOS\space}%
\providecommand \EOS [0]{\spacefactor3000\relax}%
\providecommand \BibitemShut  [1]{\csname bibitem#1\endcsname}%
\let\auto@bib@innerbib\@empty
\bibitem [{\citenamefont {Richardson}(1922)}]{richardson1922weather}%
  \BibitemOpen
  \bibfield  {author} {\bibinfo {author} {\bibfnamefont {L.~F.}\ \bibnamefont {Richardson}},\ }\href@noop {} {\emph {\bibinfo {title} {Weather Prediction by Numerical Process}}}\ (\bibinfo  {publisher} {Cambridge university press},\ \bibinfo {year} {1922})\BibitemShut {NoStop}%
\bibitem [{\citenamefont {Kolmogorov}(1941)}]{kolmogorov1941local}%
  \BibitemOpen
  \bibfield  {author} {\bibinfo {author} {\bibfnamefont {A.~N.}\ \bibnamefont {Kolmogorov}},\ }\href@noop {} {\bibfield  {journal} {\bibinfo  {journal} {C.R. Acad. Sci. URSS}\ }\textbf {\bibinfo {volume} {30}},\ \bibinfo {pages} {301} (\bibinfo {year} {1941})}\BibitemShut {NoStop}%
\bibitem [{\citenamefont {Frisch}(1995)}]{frisch1995turbulence}%
  \BibitemOpen
  \bibfield  {author} {\bibinfo {author} {\bibfnamefont {U.}~\bibnamefont {Frisch}},\ }\href@noop {} {\emph {\bibinfo {title} {Turbulence: the legacy of {AN} {K}olmogorov}}}\ (\bibinfo  {publisher} {Cambridge University Press},\ \bibinfo {year} {1995})\BibitemShut {NoStop}%
\bibitem [{\citenamefont {Kolmogorov}(1962)}]{kolmogorov1962refinement}%
  \BibitemOpen
  \bibfield  {author} {\bibinfo {author} {\bibfnamefont {A.~N.}\ \bibnamefont {Kolmogorov}},\ }\href@noop {} {\bibfield  {journal} {\bibinfo  {journal} {J. of Fluid Mechanics}\ }\textbf {\bibinfo {volume} {13}},\ \bibinfo {pages} {82} (\bibinfo {year} {1962})}\BibitemShut {NoStop}%
\bibitem [{\citenamefont {Meneveau}\ and\ \citenamefont {Sreenivasan}(1991)}]{meneveau1991multifractal}%
  \BibitemOpen
  \bibfield  {author} {\bibinfo {author} {\bibfnamefont {C.}~\bibnamefont {Meneveau}}\ and\ \bibinfo {author} {\bibfnamefont {K.~R.}\ \bibnamefont {Sreenivasan}},\ }\href@noop {} {\bibfield  {journal} {\bibinfo  {journal} {Journal of Fluid Mechanics}\ }\textbf {\bibinfo {volume} {224}},\ \bibinfo {pages} {429} (\bibinfo {year} {1991})}\BibitemShut {NoStop}%
\bibitem [{\citenamefont {Stolovitzky}\ \emph {et~al.}(1992)\citenamefont {Stolovitzky}, \citenamefont {Kailasnath},\ and\ \citenamefont {Sreenivasan}}]{stolovitzky1992kolmogorov}%
  \BibitemOpen
  \bibfield  {author} {\bibinfo {author} {\bibfnamefont {G.}~\bibnamefont {Stolovitzky}}, \bibinfo {author} {\bibfnamefont {P.}~\bibnamefont {Kailasnath}},\ and\ \bibinfo {author} {\bibfnamefont {K.~R.}\ \bibnamefont {Sreenivasan}},\ }\href@noop {} {\bibfield  {journal} {\bibinfo  {journal} {Physical Review Letters}\ }\textbf {\bibinfo {volume} {69}},\ \bibinfo {pages} {1178} (\bibinfo {year} {1992})}\BibitemShut {NoStop}%
\bibitem [{\citenamefont {Praskovsky}(1992)}]{praskovsky1992experimental}%
  \BibitemOpen
  \bibfield  {author} {\bibinfo {author} {\bibfnamefont {A.~A.}\ \bibnamefont {Praskovsky}},\ }\href@noop {} {\bibfield  {journal} {\bibinfo  {journal} {Physics of Fluids A: Fluid Dynamics}\ }\textbf {\bibinfo {volume} {4}},\ \bibinfo {pages} {2589} (\bibinfo {year} {1992})}\BibitemShut {NoStop}%
\bibitem [{\citenamefont {Wang}\ \emph {et~al.}(1996)\citenamefont {Wang}, \citenamefont {Chen}, \citenamefont {Brasseur},\ and\ \citenamefont {Wyngaard}}]{wang1996examination}%
  \BibitemOpen
  \bibfield  {author} {\bibinfo {author} {\bibfnamefont {L.-P.}\ \bibnamefont {Wang}}, \bibinfo {author} {\bibfnamefont {S.}~\bibnamefont {Chen}}, \bibinfo {author} {\bibfnamefont {J.~G.}\ \bibnamefont {Brasseur}},\ and\ \bibinfo {author} {\bibfnamefont {J.~C.}\ \bibnamefont {Wyngaard}},\ }\href@noop {} {\bibfield  {journal} {\bibinfo  {journal} {Journal of Fluid Mechanics}\ }\textbf {\bibinfo {volume} {309}},\ \bibinfo {pages} {113} (\bibinfo {year} {1996})}\BibitemShut {NoStop}%
\bibitem [{\citenamefont {Iyer}\ \emph {et~al.}(2015)\citenamefont {Iyer}, \citenamefont {Sreenivasan},\ and\ \citenamefont {Yeung}}]{iyer2015refined}%
  \BibitemOpen
  \bibfield  {author} {\bibinfo {author} {\bibfnamefont {K.~P.}\ \bibnamefont {Iyer}}, \bibinfo {author} {\bibfnamefont {K.~R.}\ \bibnamefont {Sreenivasan}},\ and\ \bibinfo {author} {\bibfnamefont {P.~K.}\ \bibnamefont {Yeung}},\ }\href@noop {} {\bibfield  {journal} {\bibinfo  {journal} {Physical Review E}\ }\textbf {\bibinfo {volume} {92}},\ \bibinfo {pages} {063024} (\bibinfo {year} {2015})}\BibitemShut {NoStop}%
\bibitem [{\citenamefont {Yeung}\ and\ \citenamefont {Ravikumar}(2020)}]{yeung2020advancing}%
  \BibitemOpen
  \bibfield  {author} {\bibinfo {author} {\bibfnamefont {P.~K.}\ \bibnamefont {Yeung}}\ and\ \bibinfo {author} {\bibfnamefont {K.}~\bibnamefont {Ravikumar}},\ }\href@noop {} {\bibfield  {journal} {\bibinfo  {journal} {Physical Review Fluids}\ }\textbf {\bibinfo {volume} {5}},\ \bibinfo {pages} {110517} (\bibinfo {year} {2020})}\BibitemShut {NoStop}%
\bibitem [{\citenamefont {Lawson}\ \emph {et~al.}(2019)\citenamefont {Lawson}, \citenamefont {Bodenschatz}, \citenamefont {Knutsen}, \citenamefont {Dawson},\ and\ \citenamefont {Worth}}]{lawson2019direct}%
  \BibitemOpen
  \bibfield  {author} {\bibinfo {author} {\bibfnamefont {J.~M.}\ \bibnamefont {Lawson}}, \bibinfo {author} {\bibfnamefont {E.}~\bibnamefont {Bodenschatz}}, \bibinfo {author} {\bibfnamefont {A.~N.}\ \bibnamefont {Knutsen}}, \bibinfo {author} {\bibfnamefont {J.~R.}\ \bibnamefont {Dawson}},\ and\ \bibinfo {author} {\bibfnamefont {N.~A.}\ \bibnamefont {Worth}},\ }\href@noop {} {\bibfield  {journal} {\bibinfo  {journal} {Physical Review Fluids}\ }\textbf {\bibinfo {volume} {4}},\ \bibinfo {pages} {022601} (\bibinfo {year} {2019})}\BibitemShut {NoStop}%
\bibitem [{\citenamefont {Hill}(2002)}]{hill2002exact}%
  \BibitemOpen
  \bibfield  {author} {\bibinfo {author} {\bibfnamefont {R.~J.}\ \bibnamefont {Hill}},\ }\href@noop {} {\bibfield  {journal} {\bibinfo  {journal} {Journal of Fluid Mechanics}\ }\textbf {\bibinfo {volume} {468}},\ \bibinfo {pages} {317} (\bibinfo {year} {2002})}\BibitemShut {NoStop}%
\bibitem [{\citenamefont {Yasuda}\ and\ \citenamefont {Vassilicos}(2018)}]{yasuda2018spatio}%
  \BibitemOpen
  \bibfield  {author} {\bibinfo {author} {\bibfnamefont {T.}~\bibnamefont {Yasuda}}\ and\ \bibinfo {author} {\bibfnamefont {J.~C.}\ \bibnamefont {Vassilicos}},\ }\href@noop {} {\bibfield  {journal} {\bibinfo  {journal} {Journal of Fluid Mechanics}\ }\textbf {\bibinfo {volume} {853}},\ \bibinfo {pages} {235} (\bibinfo {year} {2018})}\BibitemShut {NoStop}%
\bibitem [{\citenamefont {Monin}\ and\ \citenamefont {Yaglom}(1975)}]{Monin_Yaglom_1975}%
  \BibitemOpen
  \bibfield  {author} {\bibinfo {author} {\bibfnamefont {A.~S.}\ \bibnamefont {Monin}}\ and\ \bibinfo {author} {\bibfnamefont {A.~M.}\ \bibnamefont {Yaglom}},\ }\href@noop {} {\emph {\bibinfo {title} {Statistical Fluid Mechanics: Mechanics of Turbulence}}}\ (\bibinfo  {publisher} {MIT Press},\ \bibinfo {year} {1975})\BibitemShut {NoStop}%
\bibitem [{\citenamefont {Danaila}\ \emph {et~al.}(2001)\citenamefont {Danaila}, \citenamefont {Anselmet}, \citenamefont {Zhou},\ and\ \citenamefont {Antonia}}]{danaila_anselmet_zhou_antonia_2001}%
  \BibitemOpen
  \bibfield  {author} {\bibinfo {author} {\bibfnamefont {L.}~\bibnamefont {Danaila}}, \bibinfo {author} {\bibfnamefont {F.}~\bibnamefont {Anselmet}}, \bibinfo {author} {\bibfnamefont {T.}~\bibnamefont {Zhou}},\ and\ \bibinfo {author} {\bibfnamefont {R.~A.}\ \bibnamefont {Antonia}},\ }\href@noop {} {\bibfield  {journal} {\bibinfo  {journal} {Journal of Fluid Mechanics}\ }\textbf {\bibinfo {volume} {430}},\ \bibinfo {pages} {87–109} (\bibinfo {year} {2001})}\BibitemShut {NoStop}%
\bibitem [{\citenamefont {Duchon}\ and\ \citenamefont {Robert}(2000)}]{Duchon_Robert_2000}%
  \BibitemOpen
  \bibfield  {author} {\bibinfo {author} {\bibfnamefont {J.}~\bibnamefont {Duchon}}\ and\ \bibinfo {author} {\bibfnamefont {R.}~\bibnamefont {Robert}},\ }\href@noop {} {\bibfield  {journal} {\bibinfo  {journal} {Nonlinearity}\ }\textbf {\bibinfo {volume} {13}},\ \bibinfo {pages} {249} (\bibinfo {year} {2000})}\BibitemShut {NoStop}%
\bibitem [{\citenamefont {Eyink}(2002)}]{eyink2002local}%
  \BibitemOpen
  \bibfield  {author} {\bibinfo {author} {\bibfnamefont {G.~L.}\ \bibnamefont {Eyink}},\ }\href@noop {} {\bibfield  {journal} {\bibinfo  {journal} {Nonlinearity}\ }\textbf {\bibinfo {volume} {16}},\ \bibinfo {pages} {137} (\bibinfo {year} {2002})}\BibitemShut {NoStop}%
\bibitem [{\citenamefont {Dubrulle}(2019)}]{dubrulle2019beyond}%
  \BibitemOpen
  \bibfield  {author} {\bibinfo {author} {\bibfnamefont {B.}~\bibnamefont {Dubrulle}},\ }\href@noop {} {\bibfield  {journal} {\bibinfo  {journal} {Journal of Fluid Mechanics}\ }\textbf {\bibinfo {volume} {867}},\ \bibinfo {pages} {P1} (\bibinfo {year} {2019})}\BibitemShut {NoStop}%
\bibitem [{\citenamefont {Yao}\ \emph {et~al.}(2023{\natexlab{a}})\citenamefont {Yao}, \citenamefont {Schnaubelt}, \citenamefont {Szalay}, \citenamefont {Zaki},\ and\ \citenamefont {Meneveau}}]{Yao2023comparing}%
  \BibitemOpen
  \bibfield  {author} {\bibinfo {author} {\bibfnamefont {H.}~\bibnamefont {Yao}}, \bibinfo {author} {\bibfnamefont {M.}~\bibnamefont {Schnaubelt}}, \bibinfo {author} {\bibfnamefont {A.}~\bibnamefont {Szalay}}, \bibinfo {author} {\bibfnamefont {T.}~\bibnamefont {Zaki}},\ and\ \bibinfo {author} {\bibfnamefont {C.}~\bibnamefont {Meneveau}},\ }\href@noop {} {\bibfield  {journal} {\bibinfo  {journal} {Journal Fluid Mechanis (accepted)}\ } (\bibinfo {year} {2023}{\natexlab{a}})}\BibitemShut {NoStop}%
\bibitem [{\citenamefont {Yao}\ \emph {et~al.}(2023{\natexlab{b}})\citenamefont {Yao}, \citenamefont {Zaki},\ and\ \citenamefont {Meneveau}}]{yao2023entropy}%
  \BibitemOpen
  \bibfield  {author} {\bibinfo {author} {\bibfnamefont {H.}~\bibnamefont {Yao}}, \bibinfo {author} {\bibfnamefont {T.~A.}\ \bibnamefont {Zaki}},\ and\ \bibinfo {author} {\bibfnamefont {C.}~\bibnamefont {Meneveau}},\ }\href@noop {} {\bibfield  {journal} {\bibinfo  {journal} {Journal of Fluid Mechanics}\ }\textbf {\bibinfo {volume} {973}},\ \bibinfo {pages} {R6} (\bibinfo {year} {2023}{\natexlab{b}})}\BibitemShut {NoStop}%
\bibitem [{\citenamefont {Chorin}(1991)}]{chorin1991equilibrium}%
  \BibitemOpen
  \bibfield  {author} {\bibinfo {author} {\bibfnamefont {A.~J.}\ \bibnamefont {Chorin}},\ }\href@noop {} {\bibfield  {journal} {\bibinfo  {journal} {Communications in Mathematical Physics}\ }\textbf {\bibinfo {volume} {141}},\ \bibinfo {pages} {619} (\bibinfo {year} {1991})}\BibitemShut {NoStop}%
\bibitem [{\citenamefont {Paladin}\ and\ \citenamefont {Vulpiani}(1987)}]{paladin1987anomalous}%
  \BibitemOpen
  \bibfield  {author} {\bibinfo {author} {\bibfnamefont {G.}~\bibnamefont {Paladin}}\ and\ \bibinfo {author} {\bibfnamefont {A.}~\bibnamefont {Vulpiani}},\ }\href@noop {} {\bibfield  {journal} {\bibinfo  {journal} {Physics Reports}\ }\textbf {\bibinfo {volume} {156}},\ \bibinfo {pages} {147} (\bibinfo {year} {1987})}\BibitemShut {NoStop}%
\bibitem [{\citenamefont {Vela-Mart{\'\i}n}\ and\ \citenamefont {Jim{\'e}nez}(2021)}]{vela2021entropy}%
  \BibitemOpen
  \bibfield  {author} {\bibinfo {author} {\bibfnamefont {A.}~\bibnamefont {Vela-Mart{\'\i}n}}\ and\ \bibinfo {author} {\bibfnamefont {J.}~\bibnamefont {Jim{\'e}nez}},\ }\href@noop {} {\bibfield  {journal} {\bibinfo  {journal} {Journal of Fluid Mechanics}\ }\textbf {\bibinfo {volume} {915}},\ \bibinfo {pages} {A36} (\bibinfo {year} {2021})}\BibitemShut {NoStop}%
\bibitem [{\citenamefont {Castaing}(1996)}]{castaing1996temperature}%
  \BibitemOpen
  \bibfield  {author} {\bibinfo {author} {\bibfnamefont {B.}~\bibnamefont {Castaing}},\ }\href@noop {} {\bibfield  {journal} {\bibinfo  {journal} {Journal de Physique II}\ }\textbf {\bibinfo {volume} {6}},\ \bibinfo {pages} {105} (\bibinfo {year} {1996})}\BibitemShut {NoStop}%
\bibitem [{\citenamefont {Evans}\ \emph {et~al.}(1993)\citenamefont {Evans}, \citenamefont {Cohen},\ and\ \citenamefont {Morriss}}]{evans1993probability}%
  \BibitemOpen
  \bibfield  {author} {\bibinfo {author} {\bibfnamefont {D.~J.}\ \bibnamefont {Evans}}, \bibinfo {author} {\bibfnamefont {E.~G.~D.}\ \bibnamefont {Cohen}},\ and\ \bibinfo {author} {\bibfnamefont {G.~P.}\ \bibnamefont {Morriss}},\ }\href@noop {} {\bibfield  {journal} {\bibinfo  {journal} {Physical Review Letters}\ }\textbf {\bibinfo {volume} {71}},\ \bibinfo {pages} {2401} (\bibinfo {year} {1993})}\BibitemShut {NoStop}%
\bibitem [{\citenamefont {Gallavotti}\ and\ \citenamefont {Cohen}(1995)}]{gallavotti1995dynamical}%
  \BibitemOpen
  \bibfield  {author} {\bibinfo {author} {\bibfnamefont {G.}~\bibnamefont {Gallavotti}}\ and\ \bibinfo {author} {\bibfnamefont {E.~G.~D.}\ \bibnamefont {Cohen}},\ }\href@noop {} {\bibfield  {journal} {\bibinfo  {journal} {Physical Review Letters}\ }\textbf {\bibinfo {volume} {74}},\ \bibinfo {pages} {2694} (\bibinfo {year} {1995})}\BibitemShut {NoStop}%
\bibitem [{\citenamefont {Fuchs}\ \emph {et~al.}(2020)\citenamefont {Fuchs}, \citenamefont {Queir{\'o}s}, \citenamefont {Lind}, \citenamefont {Girard}, \citenamefont {Bouchet}, \citenamefont {W{\"a}chter},\ and\ \citenamefont {Peinke}}]{fuchs2020small}%
  \BibitemOpen
  \bibfield  {author} {\bibinfo {author} {\bibfnamefont {A.}~\bibnamefont {Fuchs}}, \bibinfo {author} {\bibfnamefont {S.~M.~D.}\ \bibnamefont {Queir{\'o}s}}, \bibinfo {author} {\bibfnamefont {P.~G.}\ \bibnamefont {Lind}}, \bibinfo {author} {\bibfnamefont {A.}~\bibnamefont {Girard}}, \bibinfo {author} {\bibfnamefont {F.}~\bibnamefont {Bouchet}}, \bibinfo {author} {\bibfnamefont {M.}~\bibnamefont {W{\"a}chter}},\ and\ \bibinfo {author} {\bibfnamefont {J.}~\bibnamefont {Peinke}},\ }\href@noop {} {\bibfield  {journal} {\bibinfo  {journal} {Physical Review Fluids}\ }\textbf {\bibinfo {volume} {5}},\ \bibinfo {pages} {034602} (\bibinfo {year} {2020})}\BibitemShut {NoStop}%
\bibitem [{\citenamefont {Zonta}\ and\ \citenamefont {Chibbaro}(2016)}]{zonta2016entropy}%
  \BibitemOpen
  \bibfield  {author} {\bibinfo {author} {\bibfnamefont {F.}~\bibnamefont {Zonta}}\ and\ \bibinfo {author} {\bibfnamefont {S.}~\bibnamefont {Chibbaro}},\ }\href@noop {} {\bibfield  {journal} {\bibinfo  {journal} {Europhysics Letters}\ }\textbf {\bibinfo {volume} {114}},\ \bibinfo {pages} {50011} (\bibinfo {year} {2016})}\BibitemShut {NoStop}%
\bibitem [{\citenamefont {Porporato}\ \emph {et~al.}(2020)\citenamefont {Porporato}, \citenamefont {Hooshyar}, \citenamefont {Bragg},\ and\ \citenamefont {Katul}}]{porporato2020fluctuation}%
  \BibitemOpen
  \bibfield  {author} {\bibinfo {author} {\bibfnamefont {A.}~\bibnamefont {Porporato}}, \bibinfo {author} {\bibfnamefont {M.}~\bibnamefont {Hooshyar}}, \bibinfo {author} {\bibfnamefont {A.~D.}\ \bibnamefont {Bragg}},\ and\ \bibinfo {author} {\bibfnamefont {G.}~\bibnamefont {Katul}},\ }\href@noop {} {\bibfield  {journal} {\bibinfo  {journal} {Proc. of the Royal Society A}\ }\textbf {\bibinfo {volume} {476}},\ \bibinfo {pages} {20200468} (\bibinfo {year} {2020})}\BibitemShut {NoStop}%
\bibitem [{\citenamefont {Li}\ \emph {et~al.}(2008)\citenamefont {Li}, \citenamefont {Perlman}, \citenamefont {Wan}, \citenamefont {Yang}, \citenamefont {Meneveau}, \citenamefont {Burns}, \citenamefont {Chen}, \citenamefont {Szalay},\ and\ \citenamefont {Eyink}}]{li2008public}%
  \BibitemOpen
  \bibfield  {author} {\bibinfo {author} {\bibfnamefont {Y.}~\bibnamefont {Li}}, \bibinfo {author} {\bibfnamefont {E.}~\bibnamefont {Perlman}}, \bibinfo {author} {\bibfnamefont {M.}~\bibnamefont {Wan}}, \bibinfo {author} {\bibfnamefont {Y.}~\bibnamefont {Yang}}, \bibinfo {author} {\bibfnamefont {C.}~\bibnamefont {Meneveau}}, \bibinfo {author} {\bibfnamefont {R.}~\bibnamefont {Burns}}, \bibinfo {author} {\bibfnamefont {S.}~\bibnamefont {Chen}}, \bibinfo {author} {\bibfnamefont {A.}~\bibnamefont {Szalay}},\ and\ \bibinfo {author} {\bibfnamefont {G.}~\bibnamefont {Eyink}},\ }\href@noop {} {\bibfield  {journal} {\bibinfo  {journal} {Journal of Turbulence}\ ,\ \bibinfo {pages} {N31}} (\bibinfo {year} {2008})}\BibitemShut {NoStop}%
\bibitem [{\citenamefont {Yeung}\ \emph {et~al.}(2015)\citenamefont {Yeung}, \citenamefont {Zhai},\ and\ \citenamefont {Sreenivasan}}]{yeung2015extreme}%
  \BibitemOpen
  \bibfield  {author} {\bibinfo {author} {\bibfnamefont {P.~K.}\ \bibnamefont {Yeung}}, \bibinfo {author} {\bibfnamefont {X.}~\bibnamefont {Zhai}},\ and\ \bibinfo {author} {\bibfnamefont {K.~R.}\ \bibnamefont {Sreenivasan}},\ }\href@noop {} {\bibfield  {journal} {\bibinfo  {journal} {Proceedings of the National Academy of Sciences}\ }\textbf {\bibinfo {volume} {112}},\ \bibinfo {pages} {12633} (\bibinfo {year} {2015})}\BibitemShut {NoStop}%
\bibitem [{\citenamefont {Cardesa}\ \emph {et~al.}(2017)\citenamefont {Cardesa}, \citenamefont {Vela-Mart{\'\i}n},\ and\ \citenamefont {Jim{\'e}nez}}]{cardesa2017turbulent}%
  \BibitemOpen
  \bibfield  {author} {\bibinfo {author} {\bibfnamefont {J.~I.}\ \bibnamefont {Cardesa}}, \bibinfo {author} {\bibfnamefont {A.}~\bibnamefont {Vela-Mart{\'\i}n}},\ and\ \bibinfo {author} {\bibfnamefont {J.}~\bibnamefont {Jim{\'e}nez}},\ }\href@noop {} {\bibfield  {journal} {\bibinfo  {journal} {Science}\ }\textbf {\bibinfo {volume} {357}},\ \bibinfo {pages} {782} (\bibinfo {year} {2017})}\BibitemShut {NoStop}%
\bibitem [{\citenamefont {Gallavotti}(2020)}]{gallavotti2020ensembles}%
  \BibitemOpen
  \bibfield  {author} {\bibinfo {author} {\bibfnamefont {G.}~\bibnamefont {Gallavotti}},\ }\href@noop {} {\bibfield  {journal} {\bibinfo  {journal} {The European Physical Journal E}\ }\textbf {\bibinfo {volume} {43}},\ \bibinfo {pages} {1} (\bibinfo {year} {2020})}\BibitemShut {NoStop}%
\end{thebibliography}%

\end{document}